%Paper: astro-ph/9405049
%From: R. A. Sharipov <root@bgua.bashkiria.su>
%Date: Mon, 24 May 93 08:01:41 +0600

% Typeset by AmS-TeX, version 2.1,
% Next three lines after this comment are optional, delete percent
% sign % to activate them if you have problems in compiling:
 \input amstex
 \documentstyle{amsppt}
 \pagewidth{18cm}\pageheight{23cm}
\define\rank{\operatorname{rank}}
\define\tr{\operatorname{tr}}
\loadbold
\CenteredTagsOnSplits
\rightheadtext{Complete normality conditions\dots}
\topmatter
\title
Complete normality conditions \\
for the dynamical systems on Riemannian manifolds.
\endtitle
\author
Boldin A.Yu., Bronnikov A.A., Dmitrieva V.V., Sharipov R.A.
\endauthor
\address
Department of Mathematics, Bashkir State University,
Frunze street 32, 450074 Ufa, Russia.
\endaddress
\email
root\@bgua.bashkiria.su
\endemail
\date
October 20, 1993
\enddate
\abstract
     New additional equations for the Newtonian dynamical systems
on Riemannian manifolds are found. They supplement the previously
found weak normality conditions up to the complete normality
conditions for Newtonian dynamical systems.
\endabstract
\endtopmatter
\document
\head
1. Introduction.
\endhead
     Let $M$ be the Riemannian manifold with metric tensor
$g_{ij}$. The Newtonian dynamical system on M is the law of
motion of particles on $M$ given by the following equations
in local coordinates
$$
\dot x^i=v^i\hskip 10em
\dot v^i=\Phi^i(x^1,\dots,x^n,v^1,\dots,v^n)
\tag1.1
$$
The equations \thetag{1.1} realize the Newton's second law for
the mass point with unit mass. The concept of normality for
Newtonian dynamical systems was introduced and then investigated
in the series of papers \cite{1 - 5}. It is based on the study of
some hypersurface $S$ in $M$. For some point $P$ on $S$ we define
the initial data
$$
\bigl.x^i\bigr|_{t=0}=x^i(P) \hskip 10em
\bigl.v^i\bigr|_{t=o}=v(P) n^i(P)\tag1.2
$$
for the equations \thetag{1.1}. Here $n^i(P)$ are the components
of unit normal vector at the point $P$ and $v(P)$ in \thetag{1.2}
is some scalar function on $S$. The equations \thetag{1.1} and
initial data \thetag{1.2} determine the particle flow starting
at $t=0$ from $S$ along the normal vector $\bold n(P)$ with the
initial velocity $v(P)$. For $t>0$ these particles form another
hypersurface $S_t$ and determine the family of diffeomorphisms
$f_t: S\longrightarrow S_t$.
\definition{Definition 1} The family of diffeomorphisms
$f_t$ is called the normal shift along the trajectories of
dynamical system if the trajectories of \thetag{1.1} cross
all hypersurfaces $S_t$ along their normal vectors.
\enddefinition
\definition{Definition 2} Newtonian dynamical system \thetag{1.1}
is called the dynamical system accepting the normal shift of
hypersurfaces if for any hypersurface $S\subset M$ one can find
the function $v(P)$ on $S$ such that the family of
diffeomorphisms $f_t$ given by \thetag{1.1} and \thetag{1.2} is
the normal shift of $S$.
\enddefinition
\par
     The normality condition from definition 2 was first stated
in \cite{1} and \cite{2}. By the analysis of this condition
in \cite{3} for Euclidean case $M=\Bbb R^n$ two relatively
independent normality conditions were derived: weak normality
condition and additional normality condition. Each of them is
written in the form of the system of partial differential
equations for the functions $\Phi^i$ from \thetag{1.1}. They two
both form the sufficient condition for the normality condition
from definition 2 to be fulfilled for the dynamical system
\thetag{1.1}. \par
     In \cite{5} the weak normality condition was generalized
for the non-Euclidean case of an arbitrary Riemannian manifold
$M$. In this paper we generalize the additional normality
condition from \cite{3} for the case of Riemannian manifold and
we get the complete normality condition for \thetag{1.1} in form
of aggregate of weak and additional normality conditions. Here as
in the Euclidean case of $\Bbb R^n$ the complete normality is
sufficient but not necessary for the normality condition from the
definition 2. \par
     Since it seems difficult to obtain the conditions equivalent
to definition 2 and written in form of differential equations for
the functions $\Phi^i$ we shall replace the definition 2 by more
strict definition 3. \par
\definition{Definition 3} We say that dynamical system
\thetag{1.1} satisfies the strong normality condition if
for any hypersurface $S\subset M$, for any choice of the point
$P_0\in S$ and for any real number $v_0\neq 0$ one can find
the function $v(P)$ such that it is normalized by $v(P_0)=v_0$
and the family of diffeomorphisms defined by this function is the
normal shift of hypersurface $S$.
\enddefinition
     Note here that for to avoid the multivalued functions $v(P)$
one should always consider only connected and simply connected
hypersurfaces in definitions 2 and 3. \par
\head
2. On the expansion of the algebra of tensor fields.
\endhead
     Systems of differential equations of the form \thetag{1.1}
are usually connected with the vector fields on the manifolds.
In our case  the right hand sides of \thetag{1.1} form the vector
field on the tangent bundle $TM$ for the manifold $M$. Digressing
for a while from the particular vector field given by
\thetag{1.1} let us consider some vector $\bold V$ tangent to the
tangent bundle
$$
\bold V=X^1\frac{\partial}{\partial x^1}+
\dots + X^n\frac{\partial}{\partial x^n}+
        W^1\frac{\partial}{\partial v^1}+
\dots + W^n\frac{\partial}{\partial x^n} \tag2.1
$$
First $n$ components of the vector \thetag{2.1} form the tangent
vector $\bold X=\pi(\bold V)$ to $M$. Other components of
\thetag{2.1} after some modification form another tangent vector
$\bold Z=\rho(\bold V)$ to $M$ whose components are $Z^i=W^i+
\Gamma^i_{jk}v^k X^j$. Two linear maps $\pi$ and $\rho$ project
the vector $\bold V$ onto the pair of vectors $\bold X$ and
$\bold Z$ tangent to the manifold $M$. \par
     Projections $\pi$ and $\rho$ applied to the vector fields on
$TM$ however will not give the vector fields on $M$. They give
the elements of quite other set: the expanded algebra of tensor
fields on $M$. Exact definition of tensor field from such algebra
is the following. \par
\definition{Definition 4} Tensor field of expanded algebra is a
map argument of which is a point of tangent bundle $TM$ and the
value of which is in the tensor algebra  build over the tangent
space to $M$ at the point being the projection of argument from
$TM$ to $M$.
\enddefinition
     Expanded algebra of tensor fields is equipped with the
natural operations of tensor product and contraction. It is
also equipped with two operations of covariant differentiation
$\nabla_i$ and $\tilde\nabla_i$. The detailed discussions of
these features of expanded algebra of tensor fields can be
found in \cite{5}. \par
     Each point of $TM$ is the pair of point $P\in M$ and tangent
vector  $\bold v$  at this point (vector of velocity).  If we map
each point of $TM$ onto the corresponding vector $\bold v$ then
we get the vector field $\bold v$ of expanded algebra. From
$\bold v$ we construct the scalar field $v=|\bold v|$ of expanded
algebra being the modulus of velocity. In addition to these two
fields we define the vector field $\bold N=|\bold v|^{-1}\bold v$
that consists of unit vectors directed along the vector of
velocity $\bold v$. It is defined only at that points of $TM$
where $\bold v\neq 0$. By means of the components of $\bold N$ we
construct two projector valued fields
$$
Q^i_k=N_k N^i\hskip 10em P^i_k=(\delta^i_k-N_k N^i)\tag2.2
$$
from the expanded algebra. Various relationships with
projectors $\bold Q$ and $\bold P$ from \thetag{2.2} can be found
in \cite{5}. \par
     Replacing time derivatives of velocity in \thetag{1.1} by
its covariant time derivatives we can rewrite \thetag{1.1} as
follows
$$
\dot x^i=v^i\hskip 10em
\nabla_t v^i=F^i(x^1,\dots,x^n,v^1,\dots,v^n)
\tag2.3
$$
where $F^i=\Phi^i+\Gamma^i_{jk}v^k v^j$ are the components of
vector field $\bold F$ of expanded algebra known as a force field
of Newtonian dynamical system \thetag{2.3}. \par
\head
3. Weak and complete normality conditions.
\endhead
     For the dynamical system \thetag{2.3} we consider the Cauchy
problem with initial data \thetag{1.2} on some hypersurface $S$.
Let us choose the local coordinates $u^1,\dots,u^{n-1}$ on
$S$. Solving the above Cauchy problem we obtain the family of
trajectories of the dynamical system \thetag{2.3} on $M$
parameterized by $u^1,\dots,u^{n-1}$. Variation of these
variables defines the following vectors $\boldsymbol\tau_i$
tangent to $M$
$$
\boldsymbol\tau_i=\frac{\partial x^1}{\partial u^i}
\frac{\partial}{\partial x^1}+\dots + \frac{\partial x^n}
{\partial u^i} \frac{\partial}{\partial x^n}\tag3.1
$$
on the trajectories of \thetag{2.3}. Scalar products of
$\boldsymbol\tau_1,\dots,\boldsymbol\tau_{n-1}$ and $\bold N$ are
the scalar functions $\varphi_i=\left<\boldsymbol\tau_i, \bold N
\right> = g_{kq}\tau^k_i N^q$. These functions define the mutual
orientation of trajectories and the hypersurfaces $S_t$. \par
\definition{Definition 5} Say that Newtonian dynamical system
\thetag{2.2} satisfies the weak normality condition if each
function $\varphi_i=\varphi_i(t)$ is the solution of linear
homogeneous ordinary differential equation of second order for
any choice of parametric family of trajectories of it.
\enddefinition
     The main result of \cite{5} is that the weak normality
condition is equivalent to the following system of nonlinear
partial differential equations for the force field $\bold F$
of dynamical system
$$
\aligned
&(v^{-1}F_i+\tilde\nabla_i(F^k N_k)) P^i_k = 0 \\
&(\nabla_iF_k+\nabla_kF_i-2v^{-2}F_iF_k)N^kP^i_q+ \\
&\qquad + v^{-1}(\tilde\nabla_kF_iF^k-\tilde\nabla_kF^r
N^kN_rF_i)P^i_q=0
\endaligned\tag3.2
$$
Let the weak normality condition in form of the equations
\thetag{3.2} be fulfilled. Then for to get normality in the
sense of definitions 2 and 3 we should provide the following
initial data
$$
\bigl.\varphi_i\bigr|_{t=0}=0\hskip 10em
\bigl.\dot\varphi_i\bigr|_{t=0}=0 \tag3.3
$$
for the functions $\varphi_i$ by the proper choice
of function $v(P)=v(u^1,\dots,u^{n-1})$ in \thetag{1.2}. There
$\bold n(P)$ is the unit normal vector for the hypersurface $S$.
{}From \thetag{1.2} we obtain
$$
\bigl.\bold N\bigr|_{t=0}=\bold n(u^1,\dots,u^{n-1})\tag3.4
$$
Because of \thetag{3.4} first part of the initial conditions
\thetag{3.3} is satisfied for any choice of $v(P)$. Now let us
proceed with second part of initial conditions in \thetag{3.3}
$$
\bigl.\dot\varphi_i\bigr|_{t=0}=\bigl.\nabla_t\tau^j_i
N_j \bigr|_{t=0} + \bigl. g_{jk}\tau^k_i\nabla_t N^j
\bigr|_{t=0} \tag3.5
$$
For $\nabla_t N^j$ we use \thetag{3.15} and \thetag{3.19} from
\cite{5} and then we obtain $\nabla_t N^j = v^{-1} P^j_q F^q$.
For the covariant derivatives of the vectors \thetag{3.1} we use
\thetag{3.18} from \cite{5}
$$
\bigl.\nabla_t\tau^j_i\bigr|_{t=0} = \bigl.\frac{\partial^2 x^j}
{\partial t \partial u^i}\bigr|_{t=0} + \bigl.\Gamma^j_{pq}
\frac{\partial x^p}{\partial u^i} v^q\bigr|_{t=0}
$$
Taking into account the initial data \thetag{1.2} we may
transform this relationship into the following form
$$
\bigl.\nabla_t\tau^j_i\bigr|_{t=0} = \bigl.\frac{\partial v}
{\partial u^i} n^j \bigr|_{t=0} + \bigl. v\frac{\partial n^j}
{\partial u^i}\bigr|_{t=0} + \bigl. \Gamma^j_{pq} n^q
\tau^p_i\bigr|_{t=0} \tag3.6
$$
For the further transformations of the equations \thetag{3.6} we
should recall some facts concerning submanifolds of Riemannian
spaces
$$
\align
&\frac{\partial\tau^j_k}{\partial u^i}+\Gamma^j_{pq}\tau^q_k
\tau^p_i = \hat\Gamma^q_{ik}\tau^j_q + b_{ik} n^j \tag3.7 \\
&\frac{\partial n^j}{\partial u^i}+\Gamma^j_{pq} n^q\tau^p_i =
-b^q_i\tau^j_q \tag3.8
\endalign
$$
Here $b_{ik}$ and $b^q_i$ are components of tensor of second
quadratic form and $\hat\Gamma^q_{ik}$ are components of metric
connection on $S$ defined by the metric $\hat g_{ik}=g_{pq}
\tau^p_i\tau^q_k$ induced from $M$ to the hypersurface $S$. The
equations \thetag{3.7} and \thetag{3.8} are known as Gauss and
Weingarten formulae (see \cite{6} and \cite{7}). Comparing
\thetag{3.6} with \thetag{3.8} we get
$$
\bigl.\nabla_t\tau^j_i\bigr|_{t=0}=\bigl.\frac{\partial v}
{\partial u^i} n^j\bigr|_{t=0} - \bigl.v b^q_i\tau^j_q
\bigr|_{t=0} \tag3.9
$$
Now we substitute the above obtained formula $\nabla_tN^j =
v^{-1} P^j_qF^q$ and \thetag{3.9} into \thetag{3.5}. Then from
\thetag{3.3} we obtain the following equation for still unknown
function $v = v(u^1,\dots,u^{n-1})$
$$
\frac{\partial v}{\partial u^i} = - v^{-1} g_{jk} F^j
\tau^k_i\tag3.10
$$
Components $F^j$ of the force field in \thetag{3.10} depend on
the velocity vector therefore they contain the dependence on the
unknown function $v$ in the form of $v^p=v(u^1,\dots,u^{n-1})
n^p(u^1,\dots,u^{n-1})$. \par
     The equations \thetag{3.10} form the overdetermined system
of differential equations.  For to have common solution $v$ they
should satisfy the compatibility conditions. We obtain these
conditions when consider the following derivatives
$$
\frac{\partial^2 v}{\partial u^i\partial u^j} =
\frac{\partial}{\partial u^i}\left(-v^{-1} g_{pq} F^p
\tau^q_j\right)\tag3.11
$$
calculated according to the equations \thetag{3.10}. To make
shorter all further calculations we define covariant derivatives
$D_i$ as covariant derivatives with respect to $u^i$ given by the
formula \thetag{3.18} from \cite{5}. These covariant derivatives
are applicable to the tensor-valued functions whose domain of
definition is $S$. For the tensor fields of $M$ restricted to the
hypersurface $S$ these covariant derivatives are calculated as
$D_i=\tau^k_i\nabla_k$. Note that $D_i$ are not applicable  to
tensor fields of expanded algebra unless some lifting of $S$ from
$M$ to tangent bundle $TM$ is defined. For the force field
$\bold F$ in \thetag{3.10} and \thetag{3.11} such lifting is
given by the second part of initial conditions in \thetag{1.2}.
Therefore
$$
D_iF^p=\tau^k_i\nabla_kF^p+(n^k D_iv + v D_in^k)
\tilde\nabla_kF^p\tag3.12
$$
The results of applying $D_i$ to vector fields $\tau^j_k$ and
$n^j$ are defined by Gauss and Weingarten formulae \thetag{3.7}
and \thetag{3.8}
$$
D_i\tau^j_k=\hat\Gamma^q_{ik}\tau^j_q + b_{ik} n^j
\hskip 10em D_in^j=-b^q_i\tau^j_q \tag3.13
$$
So the derivatives $D_i$ establish the link between inner geometry
of $S$ and the geometry of outer space $M$ itself. \par
     The derivatives $D_iv$ for the scalar function $v$ on $S$ are
given by the equations \thetag{3.10} the compatibility condition
for which we are going to derive now. Let's combine \thetag{3.12}
and \thetag{3.13}
$$
\aligned
&\frac{\partial^2 v}{\partial u^i\partial u^j} =
-v^{-3}F_p\tau^p_i F_q\tau^q_j - v^{-1} \nabla_pF_q\tau^p_i
\tau^q_j + \\
&+ v^{-2} F_p\tau^p_i\tilde\nabla_rF_qn^r\tau^q_j +
\tilde\nabla_pF_q b^r_i\tau^p_r\tau^q_j - v^{-1} \Gamma^k_{ij}
\tau^q_k F_q -v^{-1} b_{ij} n^q F_q
\endaligned\tag3.14
$$
After exchanging indices $i$ and $j$ in \thetag{3.14} we obtain
another expression for the same derivative in left hand side of
\thetag{3.14}. The difference of these two expressions should be
zero. This gives us the compatibility condition for \thetag{3.10}
in the following form
$$
\aligned
&\tau^k_i\tau^q_j\left(\frac{\nabla_qF_k-\nabla_kF_q}{v} +
N^r\frac{\tilde\nabla_rF_qF_k-\tilde\nabla_rF_kF_q}{v^2}
\right) + \\
& + b^r_i\tau^k_r\tau^q_j\tilde\nabla_kF_q -
b^r_j \tau^k_r \tau^q_i \tilde\nabla_kF_q = 0
\endaligned \tag3.15
$$
In order to simplify the equations \thetag{3.15} we need to
recall the following relationships due to \thetag{3.4}
$$
(g_{kr}\tau^r_i \hat g^{ij}) \tau^q_j = P^q_k \tag3.16
$$
By means of contracting \thetag{3.15} with the quantities
enclosed in brackets in \thetag{3.16} we obtain
$$
\aligned
&P^k_iP^q_j\left(\frac{\nabla_qF_k-\nabla_kF_q}{v} +
N^r\frac{\tilde\nabla_rF_qF_k-\tilde\nabla_rF_kF_q}{v^2}
\right) + \\
&+ H^k_i P^q_j \tilde\nabla_kF_q-H^k_jP^q_i\tilde\nabla_kF_q =0
\endaligned \tag3.17
$$
where $H^k_i$ are the components of the symmetric operator $\bold
H$ determined by the second quadratic form of hypersurface $S$
$$
H^k_i=g_{ir}\tau^r_q\hat g^{qj} b^p_j\tau^k_p \tag3.18
$$
Symmetric operator $\bold H$ with the matrix \thetag{3.18}
satisfies the following relationships \par
$$
\bold H\bold P = \bold P\bold H =\bold H \hskip 10em
\rank(\bold H)\le n-1 \tag3.19
$$
\proclaim{Lemma 1} For any point $P$ in $M$ and any operator
$\bold H$ in tangent space to $M$ at this point satisfying the
properties \thetag{3.19} one can find the hypersurface $S$
passing trough the point $P$ such that the matrix of the operator
$\bold H$ is given by formula \thetag{3.18} at this point.
\endproclaim
     Proof of this lemma technical. We omit it noting only that
one can choose $S$ in the class of quadrics for some local
coordinates $x^1,\dots,x^n$ in the neighborhood of $P$. \par
     Lemma 1 shows that the matrix of the operator $\bold H$ in
\thetag{3.17} is rather unrestricted. This let us replace
\thetag{3.17} by two separate equations of the following form
$$
\aligned
&P^k_iP^q_j\left(\nabla_qF_k-\nabla_kF_q +
N^r\frac{\tilde\nabla_rF_qF_k-\tilde\nabla_rF_kF_q}{v}
\right) = 0 \\
&H^k_i P^q_j \tilde\nabla_kF_q-H^k_jP^q_i\tilde\nabla_kF_q =0
\endaligned \tag3.20
$$
Now consider the operator $\bold K$ with the matrix $K^j_i=
P^k_i\tilde\nabla_kF_q P^q_r g^{rj}$. Its properties are similar
to \thetag{3.19} i.e.
$$
\bold K\bold P=\bold P\bold K=\bold K \hskip 10em
\rank(\bold K)\le n-1 \tag3.21
$$
Second equation \thetag{3.20} then means that the product
$\bold K\bold H$ is symmetric operator. Because of lemma 1 the
operator $\bold H$ is arbitrary symmetric operator. Taking
$\bold H = \bold P$ and using \thetag{3.21} we get that $\bold K$
is also the symmetric operator. The product of two symmetric
operators $\bold K\bold H$ is symmetric if and only if they are
commutating. Thus we have $\bold K\bold H=\bold H\bold K$ for
arbitrary symmetric operator $\bold H$ satisfying the
relationships \thetag{3.19}. This is possible only if $\bold K$
is proportional to $\bold P$ with some scalar factor $\bold K =
\lambda \bold P$. Scalar factor $\lambda$ is easily calculated
as $\lambda = \tr(\bold K)/\tr(\bold P)$. Now we can write
$$
\aligned
&P^k_iP^q_j\left(N^r\frac{\tilde\nabla_rF_k}{v} F_q -
\nabla_qF_k\right) = P^k_iP^q_j\left(N^r\frac{\tilde\nabla_r
F_q}{v} F_k - \nabla_kF_q\right) \\
&P^k_i\tilde\nabla_kF^q P^j_q = \frac{P^k_r\tilde\nabla_kF^q
P^r_q}{n-1} P^j_i
\endaligned\tag3.22
$$
excluding the matrix $\bold H$ from \thetag{3.20} at all. The
equations \thetag{3.22} just above combined with \thetag{3.2}
form the complete normality conditions which are sufficient for
the definitions 2 and 3 to be fulfilled for the dynamical
system \thetag{2.2}.\par
     Authors thank  the Russian  Fund  for Fundamental Researches
and the Fund ''Universities of Russia`` for financial support.
One of the authors Sharipov R.A. is grateful to International
Scientific Fund of Soros for grant of 1993 amounted in 500\$.

\enddocument